\documentclass[twocolumn,superscriptaddress,floatfix, showpacs, amsmath, amssymb, aps, prl]{revtex4-2}
\usepackage[backref=none,bookmarksnumbered=true,bookmarks=true,  	bookmarksopen=true,colorlinks=true,citecolor=blue,linkcolor=blue, 	anchorcolor=green,urlcolor=blue,unicode=false]{hyperref}

\usepackage{graphicx}
\usepackage{float}
\usepackage{dcolumn}
\usepackage{bm}
\usepackage{hyperref}
\usepackage{chemformula}
\usepackage{siunitx}
\usepackage{verbatim}
\usepackage[T1]{fontenc}
\usepackage[utf8]{inputenc}

\usepackage{times}
\usepackage[symbol]{footmisc}

\begin{document}

\author{Vibhuti N. Rai}
\email{vibhuti.rai@fu-berlin.de}
\affiliation{Fachbereich Physik, Freie Universität Berlin, 14195 Berlin, Germany}
\author{Junyoung Sim}
\affiliation{Fachbereich Physik, Freie Universität Berlin, 14195 Berlin, Germany}
\author{Florian Faaber}
\affiliation{Fachbereich Physik, Freie Universität Berlin, 14195 Berlin, Germany}
\author{Nils Bogdanoff}
\affiliation{Fachbereich Physik, Freie Universität Berlin, 14195 Berlin, Germany}
\author{Sergey Trishin}
\affiliation{Fachbereich Physik, Freie Universität Berlin, 14195 Berlin, Germany}
\author{Paul Wiechers}
\affiliation{Fachbereich Physik, Freie Universität Berlin, 14195 Berlin, Germany}
\author{Tom S. Seifert}
\affiliation{Fachbereich Physik, Freie Universität Berlin, 14195 Berlin, Germany}
\author{Tobias Kampfrath}
\affiliation{Fachbereich Physik and Halle--Berlin--Regensburg Cluster of Excellence CCE, Freie Universit\"at Berlin, Arnimallee 14, 14195 Berlin, Germany}
\author{Christian Lotze}
\affiliation{Fachbereich Physik, Freie Universität Berlin, 14195 Berlin, Germany}
\author{Katharina J. Franke}
\affiliation{Fachbereich Physik and Halle--Berlin--Regensburg Cluster of Excellence CCE, Freie Universit\"at Berlin, Arnimallee 14, 14195 Berlin, Germany}
\title{Influence of atomic-scale defects on coherent phonon excitations by THz near fields in an STM}

\begin{abstract}
Coherent phonons describe the collective, ultrafast motion of atoms and play a central role in light-induced structural dynamics. Here, we employ terahertz scanning tunneling microscopy (THz-STM) to excite and detect coherent phonons in semiconducting 2H-\ch{MoTe2} and resolve how their excitation is influenced by atomic-scale defects. In a THz pump–probe scheme, we observe long-lived oscillatory signals that we assign to out-of-plane breathing and in-plane shear modes, which are both dipole forbidden in the bulk. Remarkably, the relative excitation strength of these modes varies near defects, indicating that tip-induced local band bending modulates the coupling to the THz field. This defect-tunable coupling 
offers new opportunities to control selective excitation of vibrational modes at the nanoscale.
\end{abstract}

\maketitle
\newpage
\section{Introduction}
The ultrafast motion of atoms governs many fundamental properties of materials, such as heat, charge and spin transport, owing to the coupling between the different degrees of freedom. 
Coherent phonons are particularly interesting as they are phase-locked lattice vibrations that can provide direct insight into light-induced structural changes, electron/exciton-phonon interaction, and non-linear lattice dynamics, while also offering a way to actively modify electronic properties. \cite{hase_dynamics_2002,jeong_coherent_2016,rossi_anisotropic_2015,zhang_coherent_2020,sie_ultrafast_2019,he_coherent_2016,li_coherent_2023}.
The role of coherent lattice dynamics becomes especially important for engineering nanoscale devices based on (quasi) two-dimensional systems such as transition-metal dichalcogenides (TMDCs). Their pronounced light-matter interactions enhance coherent phonon generation and enable ultrafast structural phase transitions by optical and terahertz (THz) pulses. 

A major advantage of the THz regime is that lattice dynamics can be driven resonantly and efficiently without substantial heating \cite{kusaba_terahertz_2024,jin_enhanced_2025,zhou_terahertz_2021,jelic_terahertz_2024,shi_intrinsic_2023,roelcke_ultrafast_2024}, enabling access to non-thermal pathways for controlling structural and electronic properties. Driving by THz pulses is intriguingly versatile because it allows for the excitation of different phonon modes via various direct and indirect mechanisms \cite{juraschek_sum-frequency_2018,giorgianni_terahertz_2022,handa_terahertz_2024,maehrlein_terahertz_2017}. 
Ideally, one would like to selectively address certain modes by the THz pulses, potentially allowing for tuning material properties and enhancing the efficiency of driving phase transitions. However, unavoidable intrinsic defects may interact with the phonon modes and modify their properties \cite{Hase2010}. To understand the interaction of phonons with single defects requires a technique with both atomic-scale spatial and ultrafast temporal resolution. Furthermore, understanding the role of defects on phonon dynamics at the atomic scale highlights a pathway to engineer phonon excitations by selectively introducing defects into nanoscale materials.

\begin{figure*}
\centering
\includegraphics{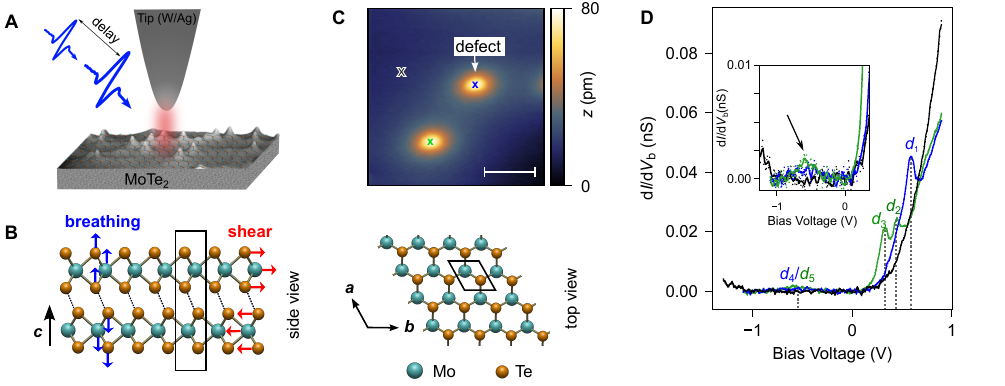}
\caption{\label{fig:MoTe2_structure}\textbf{STM characterization of 2H-\ch{MoTe2}.} \textbf{(A)} Sketch illustrating the coupling of THz pulses into the STM junction (not to scale). \textbf{(B)} Side and top view of the \ch{MoTe2} in a 2H-phase (not to scale). In-plane ($E^{2}_{2g}$) and out-of-plane ($B^{2}_{2g}$) vibrational motions are shown with red and blue arrows, respectively. \textbf{(C)} STM topography of the in-situ cleaved 2H-\ch{MoTe2}. Tunneling parameters for recording the topography are $V_\mathrm{b}$ = 1.3\,V, and $I$ = 20\,pA. The scale bar is 5\,nm. \textbf{(D)} d$I$/d$V_\mathrm{b}$ spectra recorded on the pristine surface (in black) and on two defects (in blue and green) (feedback opened at $V_\mathrm{b}$ = 0.9\,V, $I$ = 20\,pA and $V_\mathrm{mod}$ = 13\,mV). The tip positions are marked with colored crosses (black, blue, and green) in \textbf{C}. Positions of the defect states ($d_\mathrm{1}$ to $d_\mathrm{5}$) for the two defects are indicated by the vertical dashed lines. The inset shows a close-up view on the negative part of the gap, bringing out the deep in-gap defect state, marked with a black arrow. All defect states are labeled $d_i$}.
\end{figure*}

A breakthrough towards the resolution of atomic-scale dynamics has been achieved by the development of THz scanning tunneling microscopy (THz-STM), often referred as lightwave-driven STM (see Fig.\,\ref{fig:MoTe2_structure}A) \cite{cocker_ultrafast_2013, cocker_tracking_2016}. Here, a THz pulse is coupled into an STM junction, enabling field-enhanced time-resolved spectroscopy with atomic-scale spatial resolution in a pump-probe scheme \cite{cocker_ultrafast_2013,shigekawa_nanoscale_2014,cocker_tracking_2016,jelic_ultrafast_2017,yoshioka_tailoring_2018,yoshida_subcycle_2019,cocker_nanoscale_2021,yoshida_terahertz_2021,jelic_terahertz_2024,sheng_terahertz_2024,kimura_ultrafast_2025}. 
In essence, a phase-stable, single-cycle THz pulse adds an ultrafast oscillating bias voltage to the static DC bias voltage applied between the tip and the sample. This pulse leads to a transient change in the conductance, with its temporal evolution being probed by a second time-delayed pulse. By measuring the DC response, this THz pump-probe scheme enables spatially resolved investigations of dynamical processes. Recently, this method has been used to address key questions in condensed matter physics, including charge carrier dynamics \cite{cocker_ultrafast_2013,jelic_ultrafast_2017,yoshida_subcycle_2019,Allerbeck2024}, charge-density-wave dynamics \cite{sheng_terahertz_2024}, and even phase transitions \cite{jelic_terahertz_2024}. The THz field in the STM junction has also been used to excite and probe single-molecule properties \cite{cocker_tracking_2016,wang_atomic-scale_2022,kimura_ultrafast_2025}, localized vibrational modes at atomic-scale defects \cite{roelcke_ultrafast_2024,jelic_atomic-scale_2024}, and to launch and detect acoustic phonons via their reflections at material interfaces \cite{sheng_launching_2022}.

Despite these advances, the excitation and detection of intrinsic, long-range coherent optical phonons using THz-STM has remained elusive. 
Efficient coupling of a THz pulse requires a polarization that can oscillate at THz frequencies. In semiconductors, surface band bending and inefficient screening of external fields lead to charge separation that can be dynamically modulated by the THz field, resulting in a transient dipole moment that couples to the driving field.

Here, we use the near-field THz enhancement in the nanocavity of an STM tip and a semiconducting 2H-\ch{MoTe2} surface to locally excite and detect coherent phonon modes. Surface symmetry breaking and tip- and field-induced band bending allow us to address phonon modes that are forbidden in the bulk. Even more interestingly, we observe that atomic defects modify the intensity of the excited modes in the phonon spectral density. We attribute the selective enhancement to different transient dipoles in the presence of the defects due to transient charging in response to the local band bending.

\section{Results}
2H-\ch{MoTe2} has a trigonal prismatic structure, where each molybdenum (Mo) atom is sandwiched between two tellurium (Te) atoms, forming a \ch{MoTe2} layer (see Fig.\,\ref{fig:MoTe2_structure}B). These layers are stacked along the \textit{c}-axis by van der Waals interactions leading to inversion symmetry in the bulk crystal structure. Fig.\,\ref{fig:MoTe2_structure}C shows a typical STM topography of 2H-\ch{MoTe2} surface. The terraces are atomically flat but with occasional protrusions that can be attributed to intrinsic defects such as substitutional sites or vacancies (see Supplementary materials (SM) Note 1 for sample preparation) \cite{guguchia_magnetism_2018}. The differential conductance measurement (d$I$/d$V_\mathrm{b}$) on the pristine area shows a gap of around 1.2\,eV (black curve in Fig.\,\ref{fig:MoTe2_structure}D), consistent with the semiconducting nature and previous measurements \cite{lezama_surface_2014,diaz_high_2016,zhu_defects_2017}. The defect sites exhibit several in-gap states (Fig.\,\ref{fig:MoTe2_structure}D). Depending on the specific type of defects, there are one or two states close to the conduction band (CB), i.e., at $V_\mathrm{b}$ $\approx$ 0.60\,V ($d_\mathrm{1}$ in blue curve) and $V_\mathrm{b}$ $\approx$ 0.32\,V and $V_\mathrm{b}$ $\approx$ 0.44\,V ($d_\mathrm{2}$ and $d_\mathrm{3}$ in green curve), respectively. In addition, both sites show a broad defect state at $V_\mathrm{b}$ $\approx$ -0.60\,V ($d_\mathrm{4}$/$d_\mathrm{5}$) near the valence band (VB). As we will show later, the charging the defect states will have a crucial impact on the phonon excitation probability. 

\noindent\textbf{THz-STM}

\begin{figure}
\centering
\includegraphics{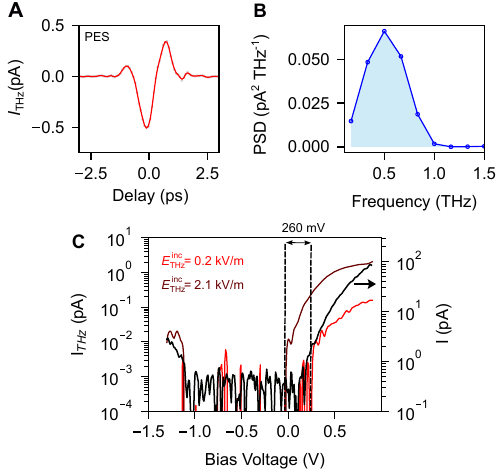}
\caption{\label{fig:pulse_shape}\textbf{THz pulse shape in the STM junction.} \textbf{(A)} Near-field pulse shape measured by photoemission sampling by a Ag tip in front of a 2H-\ch{MoTe2} surface. Parameters for acquiring the pulse shape, $V_\mathrm{b}$ = -5.0\,V, laser repetition rate $R_\mathrm{rate}$ = 1.25\,MHz. \textbf{(B)} Power spectral density (PSD) from Fast Fourier transform of the THz pulse shown in \textbf{A}. \textbf{(C)} THz-induced tunneling current ($I_\mathrm{THz}$) as a function of $V_\mathrm{b}$ at two different $E_\mathrm{THz}^\mathrm{inc}$ (feedback opened at $V_\mathrm{b}$ = 0.9\,V and $I$ = 20\,pA). The signal is integrated for 0.5\,s at each data point at a pulse repetition rate $R_\mathrm{rate}$ = 10\,MHz. For comparison, DC current is shown (in black) without any THz field.}
\end{figure}

To investigate the phonon dynamics, the STM junction is irradiated with THz pulses (Fig.\,\ref{fig:MoTe2_structure}A). Two phase stable, single-cycle THz pulses are generated by irradiating a \ch{LiNbO3} crystal with a pair of infrared (IR) pulses of 1030\,nm wavelength and 300\,fs duration (see SM Note 2 for more details) \cite{yeh_generation_2007,hoffmann_efficient_2007,hebling_generation_2008}.
It is essential to accurately characterize the near-field THz waveform in the STM junction, as reflections and low-pass filtering can distort the pulse and obscure the true dynamics of the system \cite{muller_phase-resolved_2020,sheng_control_2024}. We first verify the generation of a single-cycle THz pulse and determine its free-space field strength using electro-optic sampling (EOS) outside the junction (for the EOS details see SM Note 2 and 3). This pulse $E^\mathrm{inc}_\mathrm{THz}$ is subsequently coupled into a junction consisting of a silver (Ag)-covered tungsten tip in front of a 2H-\ch{MoTe2} sample for in-junction characterization. The generated and coupled THz pulses are linearly polarized along the tip axis. To probe the pulse shape within the junction, we use a second-harmonic pulse (515\,nm) of the IR laser to induce photoemission, which is subsequently modulated by the time-delayed THz field $E^\mathrm{inc}_\mathrm{THz}$. For these measurements the tip is held several micrometers away from the surface (for details see SM Note 3). The signal confirms the single-cycle nature of the THz pulse as shown in Fig.\,\ref{fig:pulse_shape}A and Fig.\,S3 \cite{yoshida_subcycle_2019,muller_phase-resolved_2020,martin_sabanes_femtosecond_2022,ammerman_lightwave-driven_2021,wimmer_terahertz_2014}.  
The Fourier transform of the signal shows the bandwidth of the pulse with a central frequency of $\approx$ 0.5\,THz and a strong spectral-weight cut off above $\approx$ 1.0\,THz (Fig.\,\ref{fig:pulse_shape}B). 

With the THz pulse shape and free-space strength $E^\mathrm{inc}_\mathrm{THz}$ characterized, we now use it as a transient bias voltage ($V_\mathrm{THz}$) for ultrafast measurements in the tunneling regime. Due to the field enhancement in the Ag--\ch{MoTe2} junction, the amplitude of $V_\mathrm{THz}$ must first be calibrated. In the presence of the THz pulse, the transient bias voltage $V_\mathrm{THz}(t)$ adds to the applied DC bias $V_\mathrm{b}$, such that the total voltage at the junction is $V(t)$ = $V_\mathrm{b}$ + $V_\mathrm{THz}(t)$. The resulting THz-induced current $I_\mathrm{THz} = R_\mathrm{rate} \times \int i_\mathrm{THz}(t) dt$ is measured as a modulation of the DC tunneling current with a lock-in amplifier (because the transimpedance amplifier cannot follow the sub-picosecond transient current $i_\mathrm{THz}(t)$, for details see SM Note 5). The amplitude of the transient voltage $V_\mathrm{THz}(t)$ is tuned by changing $E_\mathrm{THz}^\mathrm{inc}$, using a pair of wire grid polarizers (for details, see SM Note 2). In the absence of THz excitation ($E_\mathrm{THz}^\mathrm{inc}$ = 0), no additional current $I_\mathrm{THz}$ is detected, and only the static $I(V_\mathrm{b})$ is measured (black trace in Fig. \ref{fig:pulse_shape}C). At $E_\mathrm{THz}^\mathrm{inc}= 0.2$\,kV/m, a small induced current appears (red trace), with an onset close to that of the static $I(V_\mathrm{b})$. Increasing the far-field $E_\mathrm{THz}^\mathrm{inc}$ to 2.1\,kV/m shifts the onset of $I_\mathrm{THz}$ below the static conduction band minimum, corresponding to a  transient bias voltage $V_\mathrm{THz}\approx$ 260\,mV (Fig.\,\ref{fig:pulse_shape}C). In the following, we use this amplitude for the probe pulse, while the pump pulse is of 325\,mV. These moderate field strengths are chosen to avoid strong modification of the poorly screened electronic band structure of the semiconducting MoTe$_2$  \cite{feenstra_prospective_2009, wijnheijmer_single_2011,jelic_atomic-scale_2024}, while still yielding significant signals and efficient phonon coupling.

\noindent\textbf{THz response of MoTe$_2$}

\begin{figure}
\centering
\includegraphics{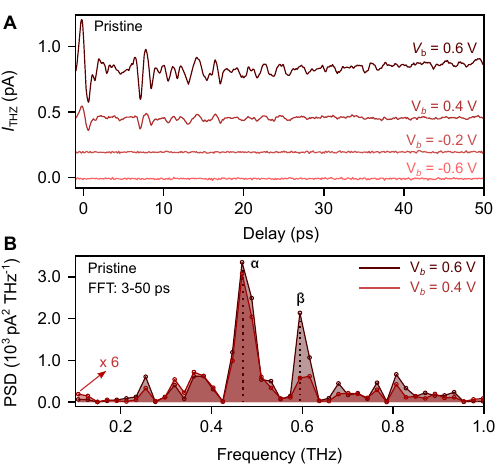}
\caption{\label{fig:FFT_compare}Terahertz pump-probe measurements on pristine 2H-\ch{MoTe2}. (a) Time traces of THz-induced tunneling current ($I_\mathrm{THz}$) on the pristine region at different bias voltages (feedback opened at $V_\mathrm{b}$ = 0.9\,V, $I$ = 10\,pA, $R_\mathrm{rate}$ = 10\,MHz). Traces are offset for clarity. (b) Power spectral density (PSD) obtained from the FFT of $I_\mathrm{THz}$ traces shown in (a). The range for the FFT is from 3 to 50 ps, which was chosen to avoid the dominant non-linearities at ultrafast timescales.
}
\end{figure}

The two THz pulses described above are time-delayed with respect to each other to excite and probe the phonons in 2H-\ch{MoTe2}. The pump pulse excites the sample by locally perturbing the substrate via its electric field, inducing an a priori unknown dynamics, which is subsequently measured by the probe pulse as a function of the time delay ($\tau$) between the two THz pulses. Fig.\,\ref{fig:FFT_compare}A shows four time traces taken at different DC bias voltages on the pristine surface, exhibiting a voltage-dependent oscillatory signal that lasts up to 50\,ps. The signal is strongly modulated over the measured time window, suggesting the presence of more than one frequency component. In contrast to the strong oscillatory signal at $V_\mathrm{b}$ = 0.6\,V, the equivalent measurements acquired with a DC bias inside the semiconducting band gap show no signal (traces at $V_\mathrm{b}$ = -0.2\,V and -0.6\,V in Fig.\,\ref{fig:FFT_compare}A). To understand the origin of the signal, we consider the effect of the two time-delayed pulses, both delivering an electric field to the junction, which is amplified at the tip apex by several orders of magnitude as has been calibrated above \cite{cocker_tracking_2016,jelic_ultrafast_2017,peller_quantitative_2021}. The first THz pulse (pump) initiates a coherent process, the dynamics of which are probed by the second pulse (probe). This leads to $I_\mathrm{THz}$, which is only measurable if the material is sufficiently conductive at the transient bias voltage, i.e., when $V_\mathrm{THz, probe}$ is sufficiently strong to reach the edge of the bandgap of 2H-\ch{MoTe2}. As a result, no signal is observed at $V_\mathrm{b}$ deep within the bandgap, consistent with our estimated amplitude $V_\mathrm{THz}$ above. Consequently, the traces at $V_\mathrm{b}$ = -0.2\,V and -0.6\,V show no induced current.

\begin{figure*}
\centering
\includegraphics{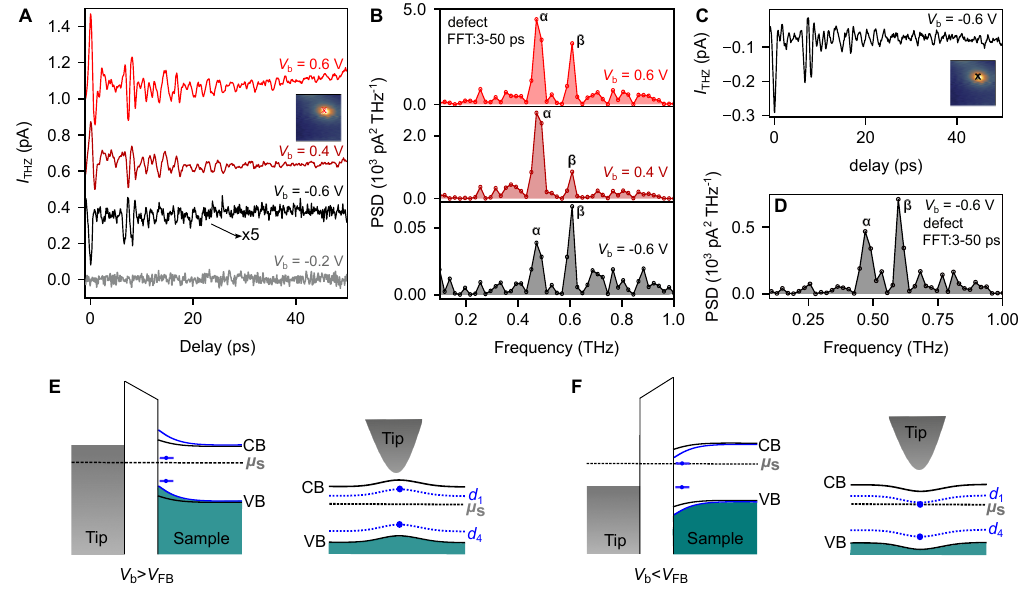}
\caption{\label{fig:defect_effect}\textbf{Effect of defects on the phonon modes.} \textbf{(A)} Time-resolved traces of $I_\mathrm{THz}$ recorded at different $V_\mathrm{b}$ (feedback opened at $V_\mathrm{b}$ = 0.9\,V, $I$ = 30\,pA, $R_\mathrm{rate}$ = 10\,MHz for all traces). Measurement positions of the time traces are marked with the (red) cross in the inset topography ($V_\mathrm{b}$\,=\,0.9\,V, $I$\,=\,4\,pA). \textbf{(B)} Power spectral density (PSD) obtained from the FFT of $I_\mathrm{THz}$ traces shown in \textbf{A}. The range for the FFT is from 3 to 50\,ps. \textbf{(C)} Time trace recorded on the same defect as in (A) but at closer tip sample distance, feedback opened at $V_\mathrm{b}$\,=\,0.6\,V, $I$\,=\,30\,pA, and \textbf{(D)} corresponding PSD. For all time-resolved traces a repetition rate $R_\mathrm{rate}$\,=\,10\,MHz is used. The signal is integrated for 0.5\,s at each data point. \textbf{(E)} \textbf{Left:} Energy diagram representing induced band bending at the surface of MoTe$_2$ at $V_\mathrm{b}$\,>\,$V_\mathrm{FB}$,  where $V_\mathrm{FB}$ is flat-band condition. Black/blue solid line represents the band bending on the pristine/defect area. \textbf{Right}: Lateral profile of the energy bands below the STM tip. Two defect states ($d_\mathrm{1}$ and $d_\mathrm{4}$) are indicated by blue dots. \textbf{(F)} \textbf{Left:} Energy diagram at $V_\mathrm{b}$\,<\,$V_\mathrm{FB}$, and \textbf{Right:} Lateral profile of the energy bands below the STM tip. In both \textbf{E} and \textbf{F}, dotted black and blue line represents the chemical potential and defect state, respectively.}
\end{figure*}

We now aim to understand the origin of the long-lasting oscillations of $I_\mathrm{THz}$ upon reaching the conduction band. We first note that the oscillatory signal cannot originate solely from the convolution of the two THz pulses with the non-linear static $I(V_\mathrm{b})$, i.e., without involving any substrate dynamics. This is confirmed by calculating the induced current $I_\mathrm{THz}$ for two time-delayed pulses (see SM Note\,5), which differs clearly from the experimental traces. Consequently, the traces must include a dynamical process within the sample. To resolve the frequencies of the induced excitations, we perform the fast Fourier transform (FFT) (Fig.\,\ref{fig:FFT_compare}B) and look at the power spectral density (PSD). For the FFT, we excluded the overlap region of the two pulses, where ultrafast processes, typically related to hot electron dynamics or other processes, such as Zener tunneling and impact ionization \cite{hirori_extraordinary_2011,lange_extremely_2014}, are beyond our time resolution. Instead, we are interested in long-lasting dynamics (for analysis details, see SM Note\,4). The FFT reveals two prominent peaks with high spectral weights at 0.48\,THz (labeled $\alpha$) and 0.6\,THz (labeled $\beta$). The low-intensity side peaks exhibit no striking variation with applied bias voltage or local defects (discussed later). The clear observation of sharp peaks signifies coherent excitations that we attribute to phonons due to their energy range being compatible with phonon modes of bulk 2H-\ch{MoTe2} \cite{froehlicher_unified_2015,grzeszczyk_raman_2016,song_physical_2016,yamamoto_strong_2014}. Group theory predicts three acoustic ($E_{1u}^{1}$ and $A_{2u}^{1}$) and fifteen optical phonon modes for bulk 2H-\ch{MoTe2}. Among the optical modes, five are doubly degenerate in-plane and another five are non-degenerate out-of-plane modes at the $\Gamma$ point, but disperse and split away from it, as reported in previous studies \cite{froehlicher_unified_2015,yamamoto_strong_2014}. With the THz pulse being confined below the tip, coupling of the transient electric field to modes close but not restricted to zero momentum, i.e., to the $\Gamma$ point, is possible. 
Based on their frequencies, we tentatively assign our observed $\alpha$ and $\beta$ modes to the in-plane shear $E^{2}_{2g}$ mode and the out-of-plane breathing $B^{2}_{2g}$ mode, respectively, though we note that prior experimental and theoretical values depend on the precise conditions prohibiting the assignment based on the precise phonon frequencies  \cite{wieting_interlayer_1980,yamamoto_strong_2014,froehlicher_unified_2015, grzeszczyk_raman_2016}.  These modes are Raman active, however, contrary to the bulk 2H-MoTe$_2$, inversion symmetry is broken at the surface, potentially rendering these also IR active.

\section*{Discussion}
\noindent\textbf{Excitation mechanism}

To understand the excitation, we note that coherent phonons can be generated through three primary mechanisms: (i) impulsive stimulated Raman scattering (ISRS) \cite{yan_impulsive_1985,dhar_time-resolved_1994}, which is a non-resonant process involving excitation of higher energy states and Stokes scattering, (ii) displacive excitation of coherent phonons (DECP) \cite{zeiger_theory_1992}, which relies on resonant electronic excitations, or (iii) direct coupling of the THz electric pump field to the dipole moment of vibrational modes \cite{huber_coherent_2015,fu_coherent_2016}. The first two mechanisms require significantly larger-energy pump pulses (or unlikely multiple photon absorption \cite{maehrlein_terahertz_2017}) for any interband transitions than provided by our THz pulses, and can therefore be excluded. This indicates that a direct coupling of the electric field to a dynamic charge distribution at the surface resonantly excites the phonons, which are well within the bandwidth of our THz pulse. Regardless of the excitation mechanism, the detection of the phonon modes relies on conductance changes, which can be induced by both the in-plane $E^{2}_{2g}$ mode and the out-of-plane breathing $B^{2}_{2g}$ mode, as both influence the local density of states below the STM tip.

To gain further insight into the excitation mechanism and coupling efficiencies, we compare the time traces at $V_\mathrm{b}$ = 0.6\,V and $V_\mathrm{b}$ = 0.4\,V in more detail (see Fig.\,\ref{fig:FFT_compare}B). Though both traces show the oscillations, their intensity is reduced at 0.4\,V compared to the larger DC voltage. The overall decrease can be attributed to the highly non-linear $I(V_\mathrm{b})$ curve, especially at the CB band onset, which effectively modifies $I_\mathrm{THz}$ of the probe pulse (also see Fig.\,S4). In contrast, the change in relative intensity of the $\alpha$ and $\beta$ mode with the $\beta$ mode showing a stronger intensity reduction than the $\alpha$ mode cannot be accounted for by the non-linearity in the $I(V_\mathrm{b})$ characteristics (see the calculated spectral response in SM Note\,5). 

To understand the energy-dependent excitation strength of the two modes and identify opportunities for its control, we investigate the THz pump-probe response at the previously identified defect sites.
To this end, we measured time traces on the defect shown above (blue curve in Fig.\,\ref{fig:MoTe2_structure}D) in different bias regimes. Figs.\,\ref{fig:defect_effect}A and B display four time traces and their corresponding FFT spectra acquired at different $V_\mathrm{b}$. To ensure a meaningful comparison, the tip--sample distance was maintained similar to that used in the measurements on the pristine surface (Figs.\,\ref{fig:FFT_compare}A). At positive bias voltages, the FFT spectra match those observed in pristine regions (Fig.\,\ref{fig:FFT_compare}A), showing the two phonon modes, $\alpha$ and $\beta$. The similarities in both the energy and relative intensities of the two modes suggest that the changes in the non-linearity of the $I(V_\mathrm{b})$ curve due to the defect states at positive $V_\mathrm{b}$ (green curve in Fig.\,\ref{fig:MoTe2_structure}D) have no major effect. However, the time trace measured at negative bias voltage displays a very different behavior. This becomes evident in the FFT signal (Fig.\,\ref{fig:defect_effect}B-D) as the relative intensity of the two modes is inverted with the $\beta$ mode being more intense than the $\alpha$ mode. 
%Instead, we \blue{consider the transient charge distribution at the surface and how this may influence the excitation efficiency}. In addition to the broken symmetry at the surface, presence of the metallic tip leads to band bending . This band bending can be further influenced by the applied bias voltage in the STM junction. Thus, we propose that the transient electric field of the THz pulse couples to the \blue{transient} surface dipole and excites coherent phonons. 

The inversion of the intensity of $\alpha$ and $\beta$ with different DC bias voltage and compared to the pristine surface cannot be explained by a simple spectral response of the two THz pulses interacting with the non-linear static $I(V_\mathrm{b})$ curve, even in the presence of the in-gap state at $V_\mathrm{b}$ = -0.6\,V (for calculated spectra see SM Note\,5). 
Instead, to explain this behavior, we consider the transient charge distribution at the surface and how this may influence the excitation efficiency. The total electric field in the junction, which consists of the tip-induced band bending $E_\mathrm{BB}$ due to the difference in the work functions of the tip and the sample, the applied DC electric field $E_\mathrm{DC}$ and the THz field $E_\mathrm{THz}(t)$: $E_\mathrm{tot}$ = $E_\mathrm{THz}(t)$ + $E_\mathrm{DC}$ + $E_\mathrm{BB}$.
We note that it is non-trivial to determine $E_\mathrm{BB}$  in an STM junction. Since the work functions of Ag (4.2-4.7\,eV, depending on crystallographic termination of the tip) and 2H-MoTe$_2$ (4.1-4.4\,eV) are nearly the same \cite{shimada_work_1994, jones_band_2022}, $E_\mathrm{BB}$ is expected to be small.

To illustrate the tip- and field-induced dipole, we present energy diagrams for two regimes at either side of the flat band ($V_\mathrm{FB}$) condition  (Fig.\,\ref{fig:defect_effect}E and F). In both cases, the applied bias voltage shifts the band relative to the flat-band condition, where the work function difference between tip and the sample is compensated by the applied $V_\mathrm{b}$.

For $V_\mathrm{b}$ > $V_\mathrm{FB}$ (Fig.\,\ref{fig:defect_effect}E), the bands bend upward, creating a charge depletion region at the MoTe${_2}$ surface and a surface dipole. The THz electric field couples to the Mo and Te ions via Coulomb interaction, accelerates them, and thus triggers coherent ionic vibrations (phonons). Because of the DC bias voltage variation, the static charge distribution is modified, which also modifies the effective charge of the ions and, thus, the phonon excitation efficiency. While we cannot make any quantitative statements on the charge distributions, this model captures the changes on the (relative) intensities of the phonon modes on the pristine surface.

In this band alignment, the defect states at positive bias voltage ($d_\mathrm{1}$) have only a little effect on the charge distribution and, consequently, on the excitation of the in-plane shear and out-of-plane breathing modes. Likewise, occupied defect states at negative bias do not alter the local charge distribution significantly as long as they remain below the chemical potential. As a result, the phonon spectra of defect and pristine areas are similar at positive bias voltage (Fig.\,\ref{fig:FFT_compare}).

In contrast, for $V_\mathrm{b}$ < $V_\mathrm{FB}$, the MoTe$_2$ band bends downwards at the interface (Fig.\,\ref{fig:defect_effect}F), again inducing a surface charge accumulation. At sufficiently strong band bending, the occupied defect state at negative bias ($d_\mathrm{1}$) crosses the chemical potential, altering the charge state, and, thereby also the transient charges that determine the phonon excitation efficiency. This model explains that the $\beta$ mode becomes more intense than the $\alpha$ mode at $V_\mathrm{b}=-0.6 \mathrm{V}$, indicating that the THz field excites the $\beta$ mode more efficiently at the defect site than on the pristine surface. 
(We note that the strong unoccupied defect states ($d_\mathrm{1}$-$d_\mathrm{3}$) are responsible for the different phonon excitation at negative bias voltage, rather than the defect states at negative bias voltage ($d_\mathrm{4}$/$d_\mathrm{5}$). For comparison of a different defect, see SM Note\,7.)

The inverse intensity ratio of the two phonon modes at the defect site compared to the pristine surface indicates that the in-plane shear and out-of-plane breathing mode respond differently to the local band bending at the defect site. The sketch of the profile of the band bending near the defect in Fig.\,\ref{fig:defect_effect}F illustrates that there is a field component parallel to the surface, enhancing the inhomogeneity of the electric field in the STM junction \cite{Krane2019}. We ascribe the qualitatively different response of the two modes to this field profile.

Different defects introduce distinct in-gap states, which modify the in-plane field component. Accordingly, measurements on various defects reveal different $\alpha$ to $\beta$ ratios (see SM Note\,7). 
The balance between the out-of-plane and in-plane components depends on factors such as the defect type, tip-sample distance, and tip geometry. Although these complexities hinder a fully quantitative description, our findings indicate that the coupling to out-of-plane breathing and in-plane shear modes by an inhomogeneous electric field \cite{jelic_atomic-scale_2024} can be locally tuned by defect-induced charge distributions.

Because the DC field scales with the tip-sample distance, the degree of band bending and, thus, the out-of-plane charge distribution varies accordingly. As shown in Fig.\ref{fig:defect_effect}C, D, reducing the tip-sample distance alters the intensity ratio of the two phonon modes albeit not as pronounced as the presence of a defect compared to the pristine surface, where there are strong changes of the transient charging.

\noindent\textbf{Conclusion}

We used THz STM to study coherent phonon excitations in 2H-\ch{MoTe2} with atomic-scale resolution. Time-resolved measurements revealed two distinct phonon modes whose relative excitation strength varies with bias voltage, but most notably locally in the presence of defects. This variation is attributed to tip-induced band bending and transient charging of defect states, which modulate the efficiency for coupling to in-plane and out-of-plane vibrational modes while leaving their frequency unchanged within our resolution.

Our results demonstrate that the local atomic structure influences coherent lattice dynamics and suggest that selective phonon excitation via local field engineering could offer a route to control material properties at the nanoscale.

\section{Acknowledgments}

We acknowledge financial support by the Deutsche Forschungsgemeinschaft (DFG, German Research Foundation) through Project No. 328545488 (CRC 227, Project No. B05).

\clearpage
\setcounter{figure}{0}
\onecolumngrid
\renewcommand{\thefigure}{S\arabic{figure}}  
\renewcommand{\figurename}{Fig.}

\section*{Supplementary Information}
\maketitle
\section{Note 1. Overview of defect density}
\begin{figure*}[ht]
\centering
\includegraphics[width = 0.4\columnwidth]{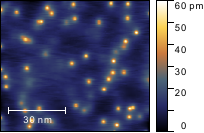}
\caption{\label{fig:topo_large} Characterization of 2H-MoTe$_2$ sample: STM topography showing a large number of defects. Scanning parameters are $V_\mathrm{b}$ = 0.7$\,$V, and $I$ = 15\,pA. }
\end{figure*}

Large scan area of the of the 2H-Mote$_2$ surface shows many bright protrusions corresponding to intrinsic defects, see Fig. \ref{fig:topo_large}. This highlights the high defect density of the sample.

\section{Note 2. Experimental setup}
\begin{figure*}[ht]
\centering
\includegraphics[width = 1\columnwidth]{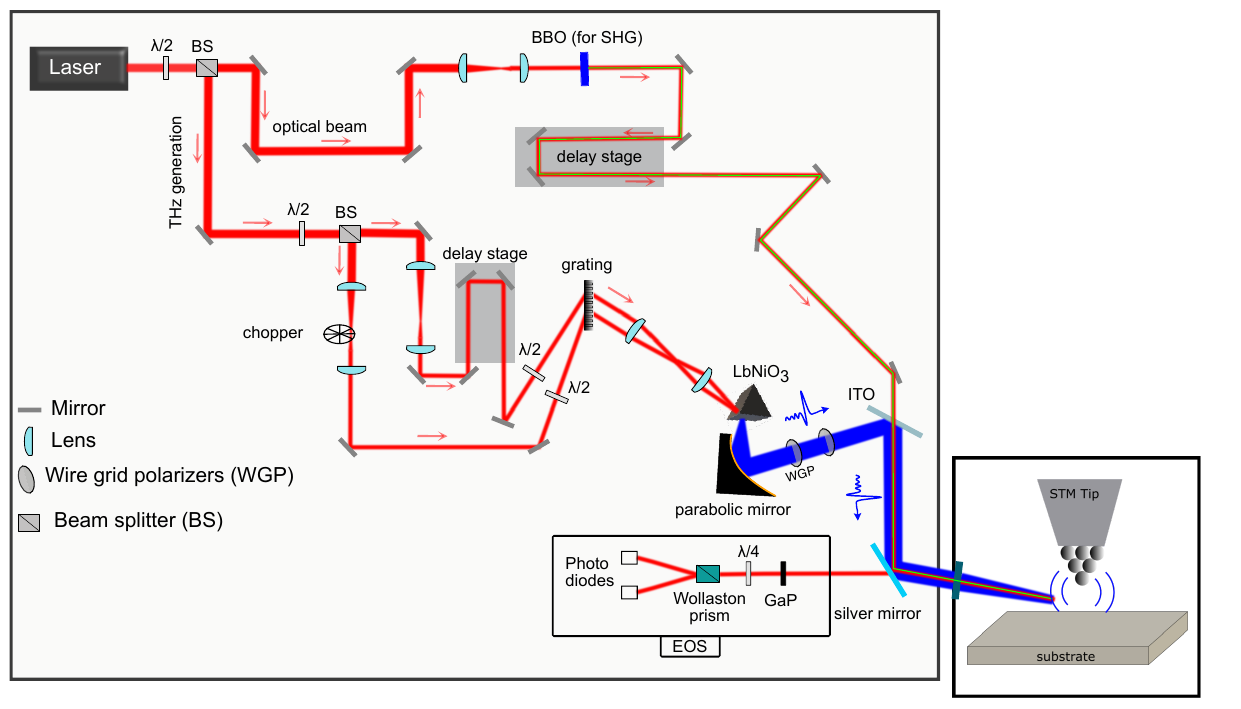}
\caption{\label{fig:setup} Scheme of the optical components and paths: The THz pulses are generated on an optical table. The high-power optical beam (1030\,nm) is marked in red, SHG (515\,nm) in green and THz in blue. Beam paths are not to scale.}
\end{figure*}

A schematic of the THz-STM setup is shown in Fig.\,\ref{fig:setup}. The optical table and STM body are mechanically decoupled. In addition, the THz beam path is enclosed in a dry-nitrogen purge box to suppress water (humidity) absorption. 

To couple the THz pulses into a STM junction, a Besocke Beetle-style STM (CreaTec Fischer and Co.) was modified. In particular, this included modification of the radiation shields.

\section{Note 3. Terahertz pulse characterization}

\begin{figure*}[ht]
\centering
\includegraphics[width = 1\columnwidth]{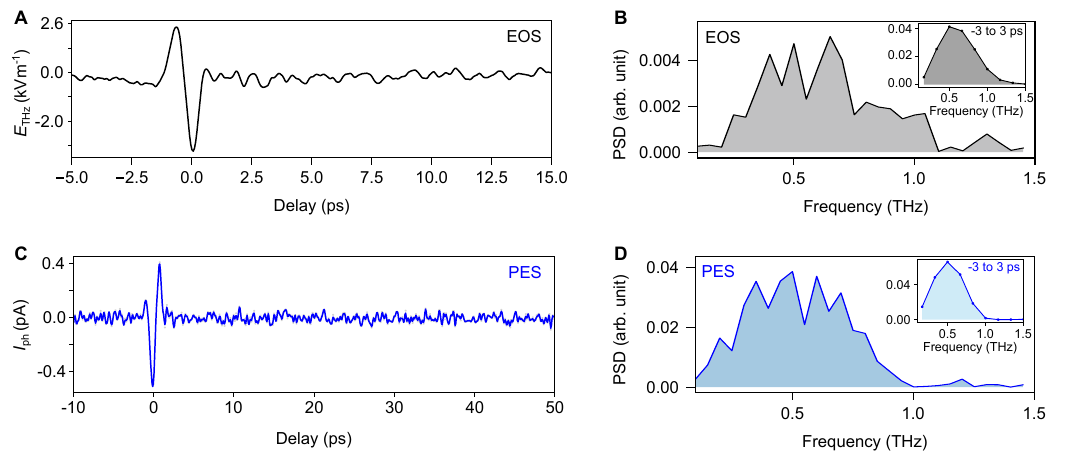}
\caption{\label{fig:spectra} \textbf{Pulse characterization: (A)} Electro-optic sampling showing the THz pulse. \textbf{(B)} FFT of the spectrum shown in \textbf{A}, taken in the temporal window of 20$\,$ps from -5$\,$ps to 15$\,$ps delay. Inset shows the FFT of the pulse in a 6$\,$ps window centered around 0$\,$ps. \textbf{(C)} Wider range of the photoemission-sampling spectrum (part of this is shown in the main text, Fig. 2A). Parameters for acquiring the pulse shape are $V_\mathrm{b}$ = -5.0$\,$V, $R_\mathrm{rate}$ = 1.25$\,$MHz. \textbf{(D)} FFT of the PES spectrum in the same spectral window as for the EOS in \textbf{B}. }
\end{figure*}

The free-space THz waveform was characterized via EOS. IR and THz beams were directed to the EOS stage or to the STM by adjusting the rotational angle of a silver mirror positioned close to the STM chamber window. For EOS, the average IR beam power was kept below 0.1$\,$W. The collinear THz and IR beams propagated through a gallium phosphide (GaP) crystal, where the THz electric field information gets encoded in the optical beam. The resulting elliptically polarized optical beam passed through a quarter-wave ($\lambda$/4) plate, converting its ellipticity into a linear polarization component. Thereafter, a Wollaston prism separates the beam into two orthogonally polarized components, and their intensity difference was measured using a balanced photodetector (BPD). By temporally sweeping the overlap between the optical main pulse and the THz pulse, the BPD output gives the temporal profile of the THz field. Fig. \ref{fig:spectra}A illustrates the far-field characteristics of the single-cycle THz pulse. The lock-in output photodetector signal was converted to electric field values using the optical properties of GaP for an IR pulse wavelength of 1030 nm and a THz central frequency of 0.5 THz, as detailed in ref. \cite{ammerman_lightwave-driven_2021}.

The temporal shape of the THz pulse near the STM tip was characterized via PES \cite{yoshida_subcycle_2019}. In this method, photoelectrons are emitted via multi-photon absorption processes in the STM junction under optical irradiation. Photoemssion was induced by 515\,nm-light pulses (generated by the second harmonic of the laser (SHG)) while maintaining a tip-sample separation of several micrometers. A DC bias voltage ($V_\mathrm{b}$) was applied across the junction to detect the photocurrent ($I_\mathrm{ph}$). The THz field modulates the potential barrier of the junction, such that by varying the temporal overlap of the SHG and THz pulses, the photocurrent is modulated. Hence, the time-delay spectra provide the temporal shape of the THz pulse that is shown in Fig. \ref{fig:spectra}C). Consistent with previous reports, the FFT of the time-delayed measurements shows that the tip acts as a low-pass filter, completely suppressing frequencies above 1\,THz, whereas the EOS shows a slightly broader frequency window (Fig. \ref{fig:spectra}B,D) \cite{muller_phase-resolved_2020}.

\section{Note 4. FFT analysis}
For identification of the phonon modes and their intensity, we calculated the fast Fourier transform (FFT) of the time-domain spectra (with Hamming window applied) (Fig. 3 and Fig. 4 in the main text) in the time window of 3-50\,ps. We exclude the pulse overlap region for two reasons: i) In the overlap region, the transient field is very high, which generates large instantaneous signals due to the strong nonlinearities. This obscures the phonon signal in the power spectral density. ii) In this regime, the dynamical processes are beyond our time resolution and not the focus of our study.

\section{Note 5. Calculated THz-induced tunneling current}

To exclude that the oscillatory behavior of $I_\mathrm{THz}$ in the time-delay measurements (Fig. 3A) is a consequence of trailing oscillations of the pulse, we calculate the convolution of the two THz pulses with the static $I(V_\mathrm{b})$ at a fixed DC bias voltage $V_\mathrm{b}$.
For the calculation, we take the measured pulse shape from PES with amplitude of 260\,mV and 325\,mV for probe and pump pulses, respectively (amplitude calibration described in the main text, Fig. 2C).
From the convolution of the two pulses with the static $I(V_\mathrm{b})$ (Fig. \ref{fig:calculation_timedelay}A), we calculate the instantaneous THz current $i_\mathrm{THz}(t)$ for different time delays ($\tau$). The superposition of the two pulses at $\tau $ = 4\,ps is shown in Fig. \ref{fig:calculation_timedelay}A (blue). 
To obtain the rectified charge $Q_\mathrm{rect}$ per pulse, $i_\mathrm{THz}(t)$ is integrated over time. The THz-induced DC current $I_\mathrm{THz}$ in Ampere is obtained by multiplication with the repetition rate. (Changes in the current in the order of picoseconds are too fast for the electronics to be picked up and are therefore averaged.) The effect of the $V_\mathrm{b}$ is incorporated by a shift of the baseline of the pulses on the $I(V_\mathrm{b})$.

In our experiments, we use an optical chopper in the IR path of the probe beam combined with a lock-in to detect the THz-induced current. 
The lock-in measures:

\begingroup
\fontsize{10}{12}\selectfont
\begin{equation}
I_{\mathrm{THz}}(\tau) = \int i_{\mathrm{THz,unblocked}}(V_\mathrm{b} + V_{2\mathrm{THz}}(\tau)) dt  - 
\int i_{\mathrm{THz,blocked}}(V_\mathrm{b} + V_\mathrm{THz}  ) dt
\label{eq:Lockin_signal_THzCurrent}
\end{equation}

Here, $i_{\mathrm{THz,unblocked}}$ denotes the case where both THz pulses are applied in the junction and induce a voltage $V_{2\mathrm{THz}}(\tau)$ that depends on the delay between the pulses $\tau$, while $i_{\mathrm{THz,blocked}}$ refers to the situation where only the unchopped pulse is present in the junction. We note that the second integral in eq. \ref{eq:Lockin_signal_THzCurrent} is independent of the time delay $\tau$ and, thus, remains constant. By subtracting this term the DC component is removed. We now compare the calculated spectra with the measured ones and their corresponding FFTs (Fig.\,\ref{fig:calculation_timedelay}B and C). It is evident that the calculations differ significantly from the experiment, indicating that the measured oscillatory behavior originates from dynamics in the junction.

\begin{figure*}[ht!]
\centering
\includegraphics{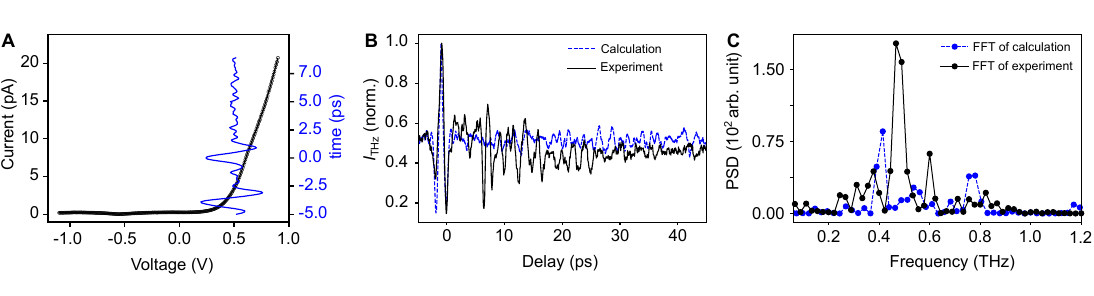}
\caption{\label{fig:calculation_timedelay} Illustration of calculations: \textbf{(A)} $I(V_\mathrm{b})$ on a defect with two pulses with $\tau = 4$\,ps superimposed at bias voltage of $V_\mathrm{b}$ = 0.5\,V. From a convolution of the two pulses with the $I(V_\mathrm{b})$ curve, the instantaneous tunneling current is calculated (blue). \textbf{(B)} Calculated (blue) and measured (black) THz-induced current with varying time delay at $V_\mathrm{b}$ = 0.4$\,$V. For the calculation, pulse shape and the amplitudes (probe is 260$\,$mV, pump is 325$\,$mV) are taken from PES and THz-induced tunneling spectroscopy measurements. The measured time trace is recorded with feedback opened at $V_\mathrm{b}$ = 0.9\,V, $I$ = 30\,pA, and $R_\mathrm{rate}$ = 10\,MHz. \textbf{(C)} The corresponding FFT is performed in the time window of 2-45\,ps, excluding the beam-overlap region.}
\end{figure*}

\begin{figure*}[ht]
\centering
\includegraphics[width = 01\columnwidth]{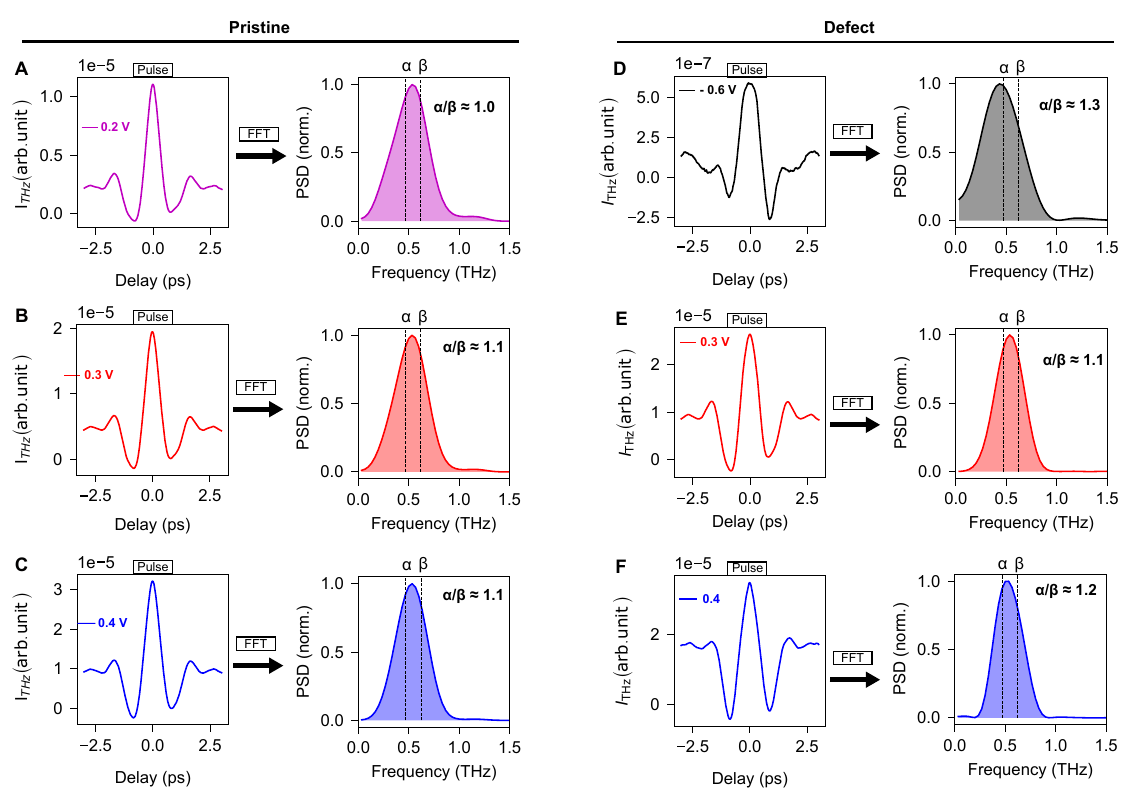}
\caption{\label{fig:calculation_bias}  Comparison of convolution of pulses in overlap region on pristine surface and defect: \textbf{(A)–(C)} \textbf{Left -} Calculated convolution of two THz pulses at different DC bias voltages ($V_\mathrm{b}$) on the pristine surface. \textbf{Right -} Corresponding FFT. \textbf{(D)–(E)} \textbf{Left -} Calculated convolution of two THz pulses at different $V_\mathrm{b}$ on a defect site. \textbf{Right -} Corresponding FFT which was performed using a Kaiser window. All calculations assume $V_\mathrm{THz}$ = 260\,mV.}
\end{figure*}

Furthermore, to exclude the possibility that the difference in the relative intensities of the two phonon modes, $\alpha$ and $\beta$, arises from the different nonlinearities of the $I(V_\mathrm{b})$ curves, we perform the calculation described above at different bias voltages on the pristine surface as well as on a defect in the range, where the two pulse overlap, and, thus, are most sensitive to the non-linearities of the $I(V_\mathrm{b})$ curve. Figure \ref{fig:calculation_bias}A-C show the rectified current (left) and the corresponding FFTs (right) in the window of -4\,ps to 2\,ps at different $V_\mathrm{b}$ on the pristine surface. The ratio of the power spectral density at the frequency of the $\alpha$ and $\beta$ (given as inset) is almost independent of bias voltage. The same behavior is observed on the defect (Fig. \ref{fig:calculation_bias}D-F). This suggests that our observation of variations in the relative intensity of the two modes must arise from the phonons being excited with different efficiency. We ascribe the different excitation efficiency of the phonon modes to changes in the local dipole moment. Most strikingly, on defects, the applied bias voltage can surpass the threshold for tip-induced charging of the defect state. In particular, we conclude that at -0.6\,V, the defect state at positive bias voltage traverses the chemical potential, leading to a charged defect. The resulting dipole moment differs significantly from the pristine area, leading to a very different coupling to the phonon modes.

\section{Note 6. Current dependence of phonon mode intensities}
We performed distance-dependent measurements on a defect site by varying the tunneling current (see Fig. \ref{fig:current_dep}A). We could only vary the current in a small range as the junction became unstable at higher currents. The overall intensities of the two modes changes monotonically as the tip-sample distance is varied (see Fig. \ref{fig:current_dep}B) without significant change in the relative intensities. Therefore, we conclude that starting with small current set points, the field strength is already sufficiently strong when starting with $V_\mathrm{b}=-0.6$\,V to charge the defect state. Larger currents and fields then only lead to an overall increase in the power spectral density of the phonon modes.

\begin{figure*}[ht]
\centering
\includegraphics{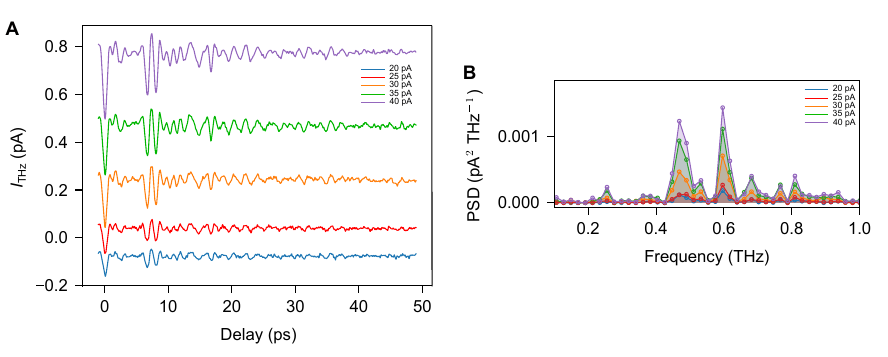}
\caption{\label{fig:current_dep} \textbf{Distance-dependent measurements: (A)} Time trace measured on the defect shown in main Fig. 4 as a function of varying set point currents. Traces are measured at $V_\mathrm{b}$ = -0.6$\,$V and laser repetition rate of $R_\mathrm{rate}$ = 10\,MHz. \textbf{(B)} FFTs of the spectra shown in A.}
\end{figure*}

\section{Note 7. Comparing different defects}
As discussed in the main text, the change in the ratio of the two phonon modes ($\alpha$/$\beta$) suggests that tip-induced band bending plays a significant role that may vary due to the presence of defects albeit with the strongest effect when defect states become charged. To further support this, we also measured time traces as a function of bias voltage on different defects with distinct defect states (see Fig. \ref{fig:cavity_response}). Figure \ref{fig:cavity_response}A and B show the $I(V_\mathrm{b})$ and d$I$/d$V_\mathrm{b}$ curves, respectively, measured on two defects and on a pristine area. The corresponding FFTs of the time traces measured at $V_\mathrm{b}$ = 0.6$\,$V show very similar total and relative intensities of the two phonon modes, consistent with the similar $I(V_\mathrm{b})$ non-linearities. As the bias voltage is decreased  ($V_\mathrm{b}$ = 0.4$\,$V), we observe a change in the relative intensities of the two modes. This further supports our hypothesis that the phonon modes are influenced by the electric field imposed by the defect and enhanced by the tip-induced band bending as the $V_\mathrm{b}$ is varied. Most striking is again the intensity inversion at $V_\mathrm{b}$ = -0.6$\,$V on the defect sites. Given the presence of defect states positive bias voltage for both types of defects, we suggest that charging of both defects takes place at negative bias voltage, leading to a strong variation in the local electric field and, thus, the coupling to the phonon modes as explained in the main text. 

\begin{figure*}[ht]
\centering
\includegraphics{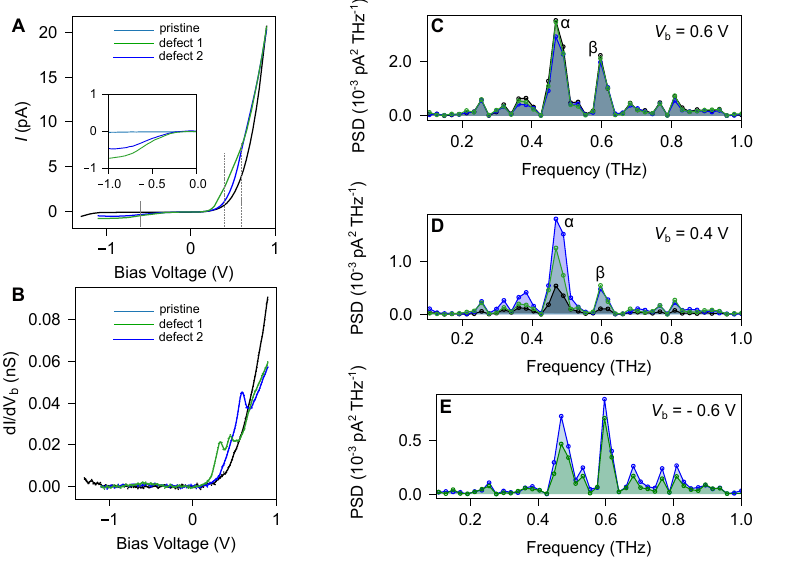}
\caption{\label{fig:cavity_response} \textbf{ Comparison of defects: (A) and (B)} $I(V_\mathrm{b})$ and d$I$/d$V_\mathrm{b}$ measured at two different defects (green and blue) and on pristine (black). \textbf{(C)}-\textbf{(E)} FFTs of the time traces measured at two defects and on pristine area at $V_\mathrm{b}$ = 0.6$\,$V, 0.4$\,$V and -0.6$\,$V, respectively. FFT of the pristine area at $V_\mathrm{b}$ = -0.6$\,$V is excluded in \textbf{E} as no oscillatory signal was observed within the band gap of the pristine MoTe$_2$.}
\end{figure*}

\end{document}